\documentclass[9pt, conference]{IEEEtran}
\IEEEoverridecommandlockouts
\usepackage{cite}
\usepackage{amsmath,amssymb,amsfonts}
\usepackage{algorithmic}
\usepackage{graphicx}
\usepackage{textcomp}
\usepackage{xcolor}
\def\BibTeX{{\rm B\kern-.05em{\sc i\kern-.025em b}\kern-.08em
    T\kern-.1667em\lower.7ex\hbox{E}\kern-.125emX}}
\usepackage{flushend}
\usepackage{balance}
\usepackage{pgfplots}
\pgfplotsset{compat=1.18}
\usepackage{tikz}
\usepackage{standalone}
\usepackage{multirow}
\usepackage{subcaption}
\usepgfplotslibrary{groupplots}
\usepackage{booktabs}
\usepackage{url}
\usepackage[hidelinks]{hyperref}

\newcommand\copyrighttext{%
  \footnotesize \textcopyright 2025 IEEE.  Personal use of this material is permitted. Permission from IEEE must be obtained for all other uses, in any current or future media, including reprinting/republishing this material for advertising or promotional purposes, creating new collective works, for resale or redistribution to servers or lists, or reuse of any copyrighted component of this work in other works.
  }
\newcommand\copyrightnotice{%
\begin{tikzpicture}[remember picture,overlay]
\node[anchor=south,yshift=10pt] at (current page.south) {\fbox{\parbox{\dimexpr\textwidth-\fboxsep-\fboxrule\relax}{\copyrighttext}}};
\end{tikzpicture}%
}

\title{Low-Resource Text-to-Speech Synthesis Using Noise-Augmented Training of ForwardTacotron}

\author{\IEEEauthorblockN{Kishor Kayyar Lakshminarayana\IEEEauthorrefmark{1},
Frank Zalkow\IEEEauthorrefmark{1},
Christian Dittmar\IEEEauthorrefmark{1}, 
Nicola Pia\IEEEauthorrefmark{1} and
{Emanu{\"e}l A.P. }{Habets}\IEEEauthorrefmark{2}}
\IEEEauthorblockA{\IEEEauthorrefmark{1}Fraunhofer Institute for Integrated Circuits (IIS), Erlangen, Germany
}
\IEEEauthorblockA{\IEEEauthorrefmark{2}{International Audio Laboratories Erlangen\textdaggerdbl,  Erlangen, Germany}\thanks{\textdaggerdbl A joint institution of the Friedrich-Alexander-Universit\"{a}t Erlangen-N\"{u}rnberg (FAU) and Fraunhofer IIS, Germany.}
}}

\begin{document}

\maketitle

\copyrightnotice
\begin{abstract}
In recent years, several text-to-speech systems have been proposed to synthesize natural speech in zero-shot, few-shot, and low-resource scenarios. However, these methods typically require training with data from many different speakers. The speech quality across the speaker set typically is diverse and imposes an upper limit on the quality achievable for the low-resource speaker. In the current work, we achieve high-quality speech synthesis using as little as five minutes of speech from the desired speaker by augmenting the low-resource speaker data with noise and employing multiple sampling techniques during training. Our method requires only four high-quality, high-resource speakers, which are easy to obtain and use in practice. Our low-complexity method achieves improved speaker similarity compared to the state-of-the-art zero-shot method HierSpeech++ and the recent low-resource method AdapterMix while maintaining comparable naturalness. Our proposed approach can also reduce the data requirements for speech synthesis for new speakers and languages.
\end{abstract}
\section{Introduction}
\label{sec:intro}
Neural non-autoregressive text-to-speech (TTS) synthesis models like FastSpeech 1 and 2 \cite{RenEtAl19_FastSpeech_NeurIPS, RenEtAl21_FastSpeech2_ICLR}, ForwardTacotron \cite{Schaefer20_ForwardTacotron_Github}, and FastPitch~\cite{Lancucki21_FastPitc_ICASSP} can synthesize nearly natural speech from text with low inference times. Many of these neural TTS models have been shown to work with multiple speakers \cite{Lancucki21_FastPitc_ICASSP, govalkar21_lightweight_ITG}. 

There is an ever-growing interest in enabling such models to perform low-resource TTS, where only a limited amount of training data for the target speaker is available. This demand has given rise to several TTS models capable of zero-shot inference like YourTTS \cite{casanova22_icml_yourtts}, HierSpeech++ \cite{lee2023_arxiv_hierspeech++}, and GZS-TV \cite{wang23c_interspeech_zero}. These models can synthesize an unseen target speaker's voice using a short speech prompt at inference time. Alternatively, neural models like Adapter-based Extension \cite{hsieh23_interspeech_adapter} and AdapterMix \cite{mehrish23_interspeech_adaptermix} can synthesize nearly natural speech using a limited amount of target speaker data at training time.

In all these low-resource TTS approaches, the base models must be trained with data from at least forty different speakers, leading to several issues. The first issue is that collecting sufficient high-quality annotated speech samples from many speakers is challenging. The second issue is that the quality of samples varies quite a bit across speakers. For example, significant variation in speech quality is seen in commonly used English datasets such as VCTK \cite{yamagishi19_vctk} and LibriTTS~\cite{zen19_libritts}. This is problematic since the training data provides an upper bound for the achievable synthesis quality. Another issue is that it is often unclear how similar the voices synthesized with massive multi-speaker TTS models are to the target speaker. In particular, it is currently under-researched whether these systems exhibit a gap in speaker similarity between the high-resource and low-resource speakers.  

Various data augmentation techniques have been proposed in the literature for achieving low-resource TTS. Some of these use powerful and pre-trained voice conversion or independent TTS models to generate augmentation data \cite{ sharma2020_strawnet_interspeech, shah2021_lowres_exp_ssw, huybrechts2021_lowres_exp_aug_icassp}, which limit the quality. Recently, Lajszczak et al. \cite{Lajszczak2022_distaug_icassp} proposed an elaborate data reordering strategy for data augmentation.  Dai et al. \cite{dai2020_noise_arxiv} and our previous work~\cite{kayyar23_eusipco_lowres} used noise augmentation for low-resource TTS. An issue with many of these data augmentation methods is that they are unsuitable in a real low-resource scenario because they require at least one hour or more of the target speaker data. Further, most of the proposed augmentation techniques are either complicated or require an additional complex pre-trained neural model.

This paper proposes a novel approach for TTS in low-resource scenarios with a low-complexity extension to ForwardTacotron \cite{Schaefer20_ForwardTacotron_Github} using data of only four high-quality high-resource speakers in addition to as little as five minutes of data from the target low-resource speaker. The proposed approach consists of three steps: 1) Splitting the low-resource samples into short segments, 2) augmenting the data using noise addition, and 3) using weighted sampling and binned sampling during the training. 

We compare our proposed model trained using different amounts of low-resource speaker data with the recently proposed zero-shot HierSpeech++ \cite{lee2023_arxiv_hierspeech++}, and the AdapterMix \cite{mehrish23_interspeech_adaptermix} models. While the proposed approach requires retraining and exhibits a slightly reduced speech naturalness compared to HierSpeech++, it achieves substantially improved speaker similarity. Moreover, increasing the low-resource training data to twenty minutes closes the gap in naturalness to the HierSpeech++ baseline. The proposed approach synthesizes speech that is significantly more natural than AdapterMix, using the same amount of low-resource (LR) data for training. Additionally, we demonstrate that the proposed approach works with only four high-quality, high-resource speakers, in contrast to the hundred or more speakers required to train the HierSpeech++ and AdapterMix models.

\section{Proposed Method}
\label{sec:method}

\subsection{Architecture}
\label{subsec:arch}
\begin{figure*}[ht]
  \centering
  \includegraphics[width=0.9\linewidth]{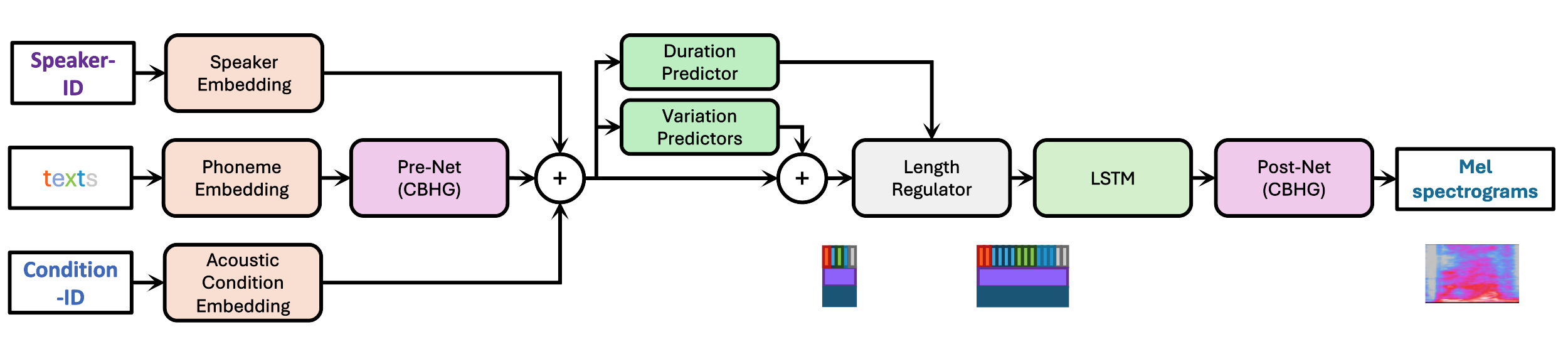}
  \caption{Block diagram of multi-speaker ForwardTacotron with the proposed extension. The $+$ symbol indicates that the tensors are concatenated. CBHG is a neural block originally proposed in Tacotron \cite{wang17n_interspeech_tacotron}. LSTM is the Long Short-Term Memory network.}
  \label{fig:forwardtacotron}
\end{figure*}

The proposed method uses two independently trained neural networks. The first network, called the acoustic model, converts text to mel-spectrogram sequences, and the second network, referred to as the neural vocoder, converts mel-spectrogram sequences to speech waveforms. In this work, we modify the acoustic model to enable low-resource TTS and use a pre-trained StyleMelGAN \cite{MustafaPF21_StyleMelGAN_ICASSP} vocoder. 

We use the non-autoregressive ForwardTacotron \cite{Schaefer20_ForwardTacotron_Github} architecture as our basis. Our model architecture is shown in Fig.~\ref{fig:forwardtacotron}. This model consists of CBHG (1-D convolution bank, highway network, and bi-directional Gated Recurrent Unit ) \cite{wang17n_interspeech_tacotron} layers as the pre- and post-nets and includes phoneme duration, pitch, energy, and voicing confidence predictors (``variation predictors'') as described by Zalkow et al.  \cite{ZalkowEtAl23_AudioLabs_Blizzard}. 

In our previous work \cite{kayyar23_eusipco_lowres}, we
suggested using three different noise augmentations in LR single-speaker auto-regressive TTS. We extend this approach to the multi-speaker non-autoregressive model using multiple augmentations with White Gaussian noise (WGN) only. We use learnable acoustic condition embeddings concatenated to the pre-net output. We use 32-dimensional vectors as these embeddings to limit the complexity addition. For a given input to the acoustic model, we select the embedding corresponding to the acoustic condition labels (cond.-ID), which we associate with our data. Each high-resource (HR) speaker is labeled with a unique cond.-ID, whereas we add two cond.-IDs for the LR speaker to enable noise augmentation. The first is the ``clean'' cond.-ID for the original low-resource data. The second is the ``noisy'' cond.-ID associated with the augmented data. For inference, only the ``clean'' cond.-ID is used.  

\subsection{Training Procedure}
\label{subsec:train}

Using a combination of HR and LR speakers to train a TTS model results in a heavily unbalanced distribution of the training samples across the speakers. Due to this imbalance, the acoustic model can learn the characteristics of the HR speakers more easily than those of the LR speaker. The few samples available for the LR speaker might be insufficient to learn its speaker-specific characteristics. Additionally, it is known that convolutional neural networks for classification do not work well when the classes in the training set are unbalanced \cite{buda18_NN_systematic}. Our problem of an unbalanced speaker distribution when training an LR multi-speaker TTS system is similar to the class imbalance problem in classification settings, especially since we also use convolutional building blocks.

To mitigate this problem, we use two strategies to increase the number of samples from the LR speaker available for training. Firstly, we split the training data into smaller segments at speech pauses using an automatic speech recognition (ASR) model like WhisperX \cite{bain2023_whisperx_interspeech}. Secondly, we create multiple noisy versions of each low-resource sample by adding WGN. Unlike our previous approach~\cite{kayyar23_eusipco_lowres}, the different additive noises are at the same power level, so the augmented samples' signal-to-noise ratio (SNR) is constant. The clean and noisy versions of the samples are distinguished using the cond.-ID described in Sec.~\ref{subsec:arch}. Although this WGN augmentation is simple, it enables high-quality synthesis in our approach. Hence, we consider more elaborate augmentation techniques unnecessary. 

If the entire training data was split into shorter segments, it could result in the model's inability to learn long-form contextual information. However, in our case, where the samples are split only for the LR speaker, the other speakers retain their long-form context. Hence, we hypothesize that the model can still learn to use long-form context from the other speakers in training. Further, due to the limited number of samples from the LR speaker in the training set, the amount of long-form context information contained in them is limited. Hence, there is only a small loss of information due to the splitting procedure.  

In addition to increasing the number of LR training items by sample splitting and noise augmentation, we employ two sampling strategies to mitigate the class imbalance problem further: weighted sampling~\cite{buda18_NN_systematic} and binned sampling. In weighted sampling, samples from the under-represented class, from the LR speaker, are drawn more often during training. 

In neural network training, the training loss is averaged across the samples in the batch. In an unbalanced training set, the content of randomly drawn batches is consequently also unbalanced. Therefore, the contribution of the LR speaker to the weight update is smaller than that of the HR speakers. We use a binned sampling strategy to increase the contribution of the LR speaker. For a binned sampling procedure, separate HR and LR bins are created for the HR and LR training samples, respectively. A training batch is selected at random, proportional to the bin sizes, from one of the bins. This strategy ensures that some batches in training consist only of LR samples, such that the training loss is only affected by these samples in this case.


\section{Experiments}
\label{sec:exp}
\setlength{\belowcaptionskip}{-10pt}
\begin{table}[]
    \caption{Details of the datasets.}
    \centering{\small
    \begin{tabular}{l l r r}
         \toprule
         Dataset & Gender & Duration  & Files \\
         \midrule
         LJSpeech \cite{ItoJohnson17_LJSpeech_ONLINE} & female & 23.5 hours & 13\,100\\
         Proprietary & male &  2.2 hours & 4400 \\
         Proprietary & female &  2.3 hours & 2700 \\
         TC-Star \cite{bonafonte06-elra_tc} & male & 5.5 hours & 4900 \\
         Hi-Fi-TTS ``92'' \cite{bakhturina21_interspeech_HiFi}
         & female & 23 hours & 34\,100 \\
         \bottomrule
    \end{tabular}
    }
    \label{tab:datasets}
\end{table}
\setlength{\belowcaptionskip}{-10pt}
\begin{table*}[t]
\caption{Objective metrics with 95\% confidence intervals across different low-resource (LR) simulations and baselines for the LR/one-shot speaker using male (TC-Star) and female (Hi-Fi-TTS-92) speakers. The bottom section has the metrics for the ablations.}
  \label{tab:eval-metrics}
  \centering{\small
  \begin{tabular}{l c c c c }
    \toprule
    \multirow{2}{*}{Model or LR subset used} & \multicolumn{2}{c}{TC-Star} & \multicolumn{2}{c}{Hi-Fi-TTS ``92''} \\
    & MCD-DTW ($\downarrow$) & cos-sim ($\uparrow$) & MCD-DTW ($\downarrow$) & cos-sim ($\uparrow$) \\
    \midrule
    ForwardTacotron (with HR datasets only) & $44.6 \pm 0.8$ & $0.65 \pm 0.009$ & $39.5 \pm 0.6$ & $0.83 \pm 0.007$ \\
    \midrule
    Proposed (20 min.) & $50.6 \pm 0.8$ & $0.56 \pm 0.009$ & $47.2 \pm 0.6$ & $0.75 \pm 0.008$\\
    Proposed (5 min.) & $53.7 \pm 0.8$ & $0.47 \pm 0.012$ & $49.6 \pm 0.6$ & $0.69 \pm 0.009$\\
    Proposed (1 min.) & $59.7 \pm 1.1$ & $0.34 \pm 0.028$ & $50.6 \pm 0.6$ & $0.60 \pm 0.009$\\
    \midrule
    HierSpeech++ & $54.9 \pm 0.6$ & $0.30 \pm 0.008$ & $52.8 \pm 0.8$ & $0.45 \pm 0.009$\\
    AdapterMix (20 min.) & $53.7 \pm 0.8$ & $0.44 \pm 0.011$ & $49.5 \pm 0.6$ & $0.64 \pm 0.007$\\
    \midrule
    Proposed (5 min. w/o binning) & $61.3 \pm 1.2$ & $0.31 \pm 0.024$ & $48.7 \pm 0.7$ & $0.69 \pm 0.011$\\
    Proposed (5 min. w/o (noise + binning)) & $61.1 \pm 1.3$ & $ 0.28 \pm 0.018$ & $53.5 \pm 1.0$ & $0.48 \pm 0.038$\\
    \bottomrule
  \end{tabular}
  }
\end{table*}

We used five high-quality speakers from multiple datasets for our experiments, listed in Table \ref{tab:datasets}. The first three speakers were used entirely for training and are considered HR speakers. As a fourth HR speaker, either Speaker 92 from the Hi-Fi-TTS dataset or the male speaker from the TC-Star dataset was used in its entirety. A subset of the data from the other speaker was used to simulate the LR training data. The LR subsets contain 20 minutes, 5 minutes, or 1 minute of speech. We use all four HR speakers and one of the LR subsets for each experiment. We always use short samples for these subsets to maximize the number of LR speaker samples available. To generate the subsets, we arrange the samples in the order of increasing duration and select samples from the beginning of this list.

The Voicebox toolbox\footnote{\url{http://www.ee.ic.ac.uk/hp/staff/dmb/voicebox/voicebox.html}} was used for the WGN addition. This toolbox determines the SNR based on the active speech power following the ITU-T P.56 recommendation \cite{itu_p56_2011}. 

In preliminary experiments, we observed that an average of around a thousand LR sentences provided training stability for our model. We ensured this minimum number of sentences through noise augmentation and our proposed training strategies described in Sec.~\ref{subsec:train}.  In the 5-minute and 20-minute subsets, we used five noise augmentations, meaning each LR sample was augmented with WGN five times at the same SNR level (20~dB), resulting in more than 1000 sentences in each case. In these cases, no weighted sampling was used. Using 1 minute of training data, 10 WGN noise augmentations and weighting by a factor of 6 were used to reach the 1000 sentence threshold. Using more noise augmentations or a higher weighting factor did not improve the performance. 

In addition to our main experiments, we also report on several side investigations (see Sec. \ref{subsec:ablation}). In particular, we repeated the experiments with multiple subsets to verify that the results were reproducible. Furthermore, we confirmed the requirement of short samples by testing our approach using a subset containing only long sentences, resulting in decreased speech quality. As an alternative to using short samples from the dataset, we also conducted an experiment where we split longer sentences into shorter segments at speech pauses using WhisperX \cite{bain2023_whisperx_interspeech}, which worked as good as using short sentences.  


\section{Results}
\label{sec:res}

We evaluate our proposed and three other speech synthesis models. The first model is ForwardTacotron \cite{Schaefer20_ForwardTacotron_Github} trained with the complete data from all the five speakers presented in Table~\ref{tab:datasets}. The second baseline is a pre-trained HierSpeech++ model \cite{lee2023_arxiv_hierspeech++}, which we use for zero-shot inference. We use the HierSpeech++ model trained with 2311 LibriTTS speakers published by the authors\footnote{\url{https://github.com/sh-lee-prml/HierSpeechpp}}. The third baseline is the AdapterMix model \cite{mehrish23_interspeech_adaptermix}.  
We pre-trained the AdapterMix model\footnote{\url{https://github.com/declare-lab/adapter-mix}} for 1800K steps using all the 109 speakers from the VCTK \cite{yamagishi19_vctk} dataset and finetuned it for 50K steps using our 20-minute subset with short sentences for the desired LR speaker (batch size 16). 
In terms of complexity, HierSpeech++ uses 108.5M parameters, AdapterMix requires 52M parameters, whereas our proposed approach uses only 43M parameters and is the least complex.

We trained our model with four HR speakers (described in Sec.~\ref{sec:exp}), and one of the  LR subsets from the TC-Star or Hi-Fi-TTS dataset, respectively, each consisting of 1 minute, 5 minutes, or 20 minutes duration. This resulted in six different training instances overall. Each instance used a batch size of 32 and was terminated after 300K training steps. 

\subsection{Objective Evaluation}
\label{subsec:obj}
\setlength{\belowcaptionskip}{-10pt}

\begin{figure*}[ht]
\centering
\begin{tikzpicture}

\definecolor{darkgray176}{RGB}{176,176,176}
\definecolor{darkslategray61}{RGB}{61,61,61}
\definecolor{forestgreen4416044}{RGB}{44,160,44}
\definecolor{lightgray204}{RGB}{204,204,204}

\begin{groupplot}[group style={group size=2 by 1, vertical sep=2cm}]
\nextgroupplot[
tick align=outside,
tick pos=left,
x grid style={darkgray176},
xmin=-0.5, xmax=4.5,
xtick style={color=black},
xtick={0,1,2,3,4},
xticklabels={HighRes,20min,5min,1min,HierSp++},
y grid style={darkgray176},
ylabel={Naturalness MOS},
ymajorgrids,
ymin=0.8, ymax=5.2,
ytick style={color=black}
]
\path [draw=blue, thick]
(axis cs:-0.1,3)
--(axis cs:0.1,3)
--(axis cs:0.1,4.25)
--(axis cs:-0.1,4.25)
--(axis cs:-0.1,3)
--cycle;
\path [draw=blue, thick]
(axis cs:0.9,3)
--(axis cs:1.1,3)
--(axis cs:1.1,4)
--(axis cs:0.9,4)
--(axis cs:0.9,3)
--cycle;
\path [draw=blue, thick]
(axis cs:1.9,2)
--(axis cs:2.1,2)
--(axis cs:2.1,4)
--(axis cs:1.9,4)
--(axis cs:1.9,2)
--cycle;
\path [draw=blue, thick]
(axis cs:2.9,1)
--(axis cs:3.1,1)
--(axis cs:3.1,2)
--(axis cs:2.9,2)
--(axis cs:2.9,1)
--cycle;
\path [draw=blue, thick]
(axis cs:3.9,3)
--(axis cs:4.1,3)
--(axis cs:4.1,5)
--(axis cs:3.9,5)
--(axis cs:3.9,3)
--cycle;
\addplot [thick, blue]
table {%
0 3
0 2
};
\addplot [thick, blue]
table {%
0 4.25
0 5
};
\addplot [thick, blue]
table {%
-0.05 2
0.05 2
};
\addplot [thick, blue]
table {%
-0.05 5
0.05 5
};
\addplot [black, mark=diamond*, mark size=2.5, mark options={solid,fill=darkslategray61}, only marks]
table {%
0 1
0 1
};
\addplot [thick, blue]
table {%
1 3
1 2
};
\addplot [thick, blue]
table {%
1 4
1 5
};
\addplot [thick, blue]
table {%
0.95 2
1.05 2
};
\addplot [thick, blue]
table {%
0.95 5
1.05 5
};
\addplot [black, mark=diamond*, mark size=2.5, mark options={solid,fill=darkslategray61}, only marks]
table {%
1 1
1 1
1 1
1 1
1 1
1 1
};
\addplot [thick, blue]
table {%
2 2
2 1
};
\addplot [thick, blue]
table {%
2 4
2 5
};
\addplot [thick, blue]
table {%
1.95 1
2.05 1
};
\addplot [thick, blue]
table {%
1.95 5
2.05 5
};
\addplot [thick, blue]
table {%
3 1
3 1
};
\addplot [thick, blue]
table {%
3 2
3 3
};
\addplot [thick, blue]
table {%
2.95 1
3.05 1
};
\addplot [thick, blue]
table {%
2.95 3
3.05 3
};
\addplot [black, mark=diamond*, mark size=2.5, mark options={solid,fill=darkslategray61}, only marks]
table {%
3 4
3 5
3 5
3 4
3 4
3 4
3 4
3 4
3 4
3 4
3 4
3 4
3 4
3 4
3 4
3 4
3 4
3 5
3 4
3 5
3 4
3 4
3 5
3 5
3 4
3 5
3 5
3 4
};
\addplot [thick, blue]
table {%
4 3
4 1
};
\addplot [thick, blue]
table {%
4 5
4 5
};
\addplot [thick, blue]
table {%
3.95 1
4.05 1
};
\addplot [thick, blue]
table {%
3.95 5
4.05 5
};
\addplot [ultra thick, red]
table {%
-0.1 4
0.1 4
};
\addplot [ultra thick, forestgreen4416044, dotted]
table {%
-0.1 3.88333333333333
0.1 3.88333333333333
};
\addplot [ultra thick, red]
table {%
0.9 4
1.1 4
};
\addplot [ultra thick, forestgreen4416044, dotted]
table {%
0.9 3.69166666666667
1.1 3.69166666666667
};
\addplot [ultra thick, red]
table {%
1.9 3
2.1 3
};
\addplot [ultra thick, forestgreen4416044, dotted]
table {%
1.9 2.87222222222222
2.1 2.87222222222222
};
\addplot [ultra thick, red]
table {%
2.9 2
3.1 2
};
\addplot [ultra thick, forestgreen4416044, dotted]
table {%
2.9 1.98055555555556
3.1 1.98055555555556
};
\addplot [ultra thick, red]
table {%
3.9 4
4.1 4
};
\addplot [ultra thick, forestgreen4416044, dotted]
table {%
3.9 3.99166666666667
4.1 3.99166666666667
};

\nextgroupplot[xshift=1cm,
legend cell align={left},
legend columns=2,
legend style={
  fill opacity=1,
  draw opacity=1,
  text opacity=1,
  at={(0.5,-0.125)},
  anchor=north west,
  draw=lightgray204
},
tick align=outside,
tick pos=left,
x grid style={darkgray176},
xmin=-0.5, xmax=4.5,
xtick style={color=black},
xtick={0,1,2,3,4},
xticklabels={HighRes,20min,5min,1min,HierSp++},
y grid style={darkgray176},
ylabel={Speaker Similarity MOS},
ymajorgrids,
ymin=0.8, ymax=5.2,
ytick style={color=black}
]
\path [draw=blue, thick]
(axis cs:-0.1,4)
--(axis cs:0.1,4)
--(axis cs:0.1,5)
--(axis cs:-0.1,5)
--(axis cs:-0.1,4)
--cycle;
\path [draw=blue, thick]
(axis cs:0.9,3)
--(axis cs:1.1,3)
--(axis cs:1.1,5)
--(axis cs:0.9,5)
--(axis cs:0.9,3)
--cycle;
\path [draw=blue, thick]
(axis cs:1.9,2)
--(axis cs:2.1,2)
--(axis cs:2.1,4)
--(axis cs:1.9,4)
--(axis cs:1.9,2)
--cycle;
\path [draw=blue, thick]
(axis cs:2.9,1)
--(axis cs:3.1,1)
--(axis cs:3.1,2)
--(axis cs:2.9,2)
--(axis cs:2.9,1)
--cycle;
\path [draw=blue, thick]
(axis cs:3.9,2)
--(axis cs:4.1,2)
--(axis cs:4.1,4)
--(axis cs:3.9,4)
--(axis cs:3.9,2)
--cycle;
\addplot [thick, blue, forget plot]
table {%
0 4
0 3
};
\addplot [thick, blue, forget plot]
table {%
0 5
0 5
};
\addplot [thick, blue, forget plot]
table {%
-0.05 3
0.05 3
};
\addplot [thick, blue, forget plot]
table {%
-0.05 5
0.05 5
};
\addplot [black, mark=diamond*, mark size=2.5, mark options={solid,fill=darkslategray61}, only marks, forget plot]
table {%
0 2
0 2
0 1
0 2
0 2
0 2
0 2
0 2
0 2
0 2
0 2
0 2
0 2
0 2
0 2
0 2
0 2
0 2
0 2
0 2
0 2
0 2
};
\addplot [thick, blue, forget plot]
table {%
1 3
1 1
};
\addplot [thick, blue, forget plot]
table {%
1 5
1 5
};
\addplot [thick, blue, forget plot]
table {%
0.95 1
1.05 1
};
\addplot [thick, blue, forget plot]
table {%
0.95 5
1.05 5
};
\addplot [thick, blue, forget plot]
table {%
2 2
2 1
};
\addplot [thick, blue, forget plot]
table {%
2 4
2 5
};
\addplot [thick, blue, forget plot]
table {%
1.95 1
2.05 1
};
\addplot [thick, blue, forget plot]
table {%
1.95 5
2.05 5
};
\addplot [thick, blue, forget plot]
table {%
3 1
3 1
};
\addplot [thick, blue, forget plot]
table {%
3 2
3 3
};
\addplot [thick, blue, forget plot]
table {%
2.95 1
3.05 1
};
\addplot [thick, blue, forget plot]
table {%
2.95 3
3.05 3
};
\addplot [black, mark=diamond*, mark size=2.5, mark options={solid,fill=darkslategray61}, only marks, forget plot]
table {%
3 5
3 5
3 5
3 5
3 4
3 4
3 4
3 5
3 4
3 4
3 4
3 4
3 4
3 4
3 4
3 4
3 4
3 4
3 5
3 4
3 5
3 5
3 4
3 4
3 4
3 4
3 4
3 4
3 4
3 5
};
\addplot [thick, blue, forget plot]
table {%
4 2
4 1
};
\addplot [thick, blue, forget plot]
table {%
4 4
4 5
};
\addplot [thick, blue, forget plot]
table {%
3.95 1
4.05 1
};
\addplot [thick, blue, forget plot]
table {%
3.95 5
4.05 5
};
\addplot [ultra thick, red, forget plot]
table {%
-0.1 5
0.1 5
};
\addplot [ultra thick, forestgreen4416044, dotted, forget plot]
table {%
-0.1 4.24
0.1 4.24
};
\addplot [ultra thick, red, forget plot]
table {%
0.9 4
1.1 4
};
\addplot [ultra thick, forestgreen4416044, dotted, forget plot]
table {%
0.9 3.89
1.1 3.89
};
\addplot [ultra thick, red, forget plot]
table {%
1.9 3
2.1 3
};
\addplot [ultra thick, forestgreen4416044, dotted, forget plot]
table {%
1.9 3.03666666666667
2.1 3.03666666666667
};
\addplot [ultra thick, red]
table {%
2.9 1
3.1 1
};
\addlegendentry{Median}
\addplot [ultra thick, forestgreen4416044, dotted, forget plot]
table {%
2.9 1.61333333333333
3.1 1.61333333333333
};
\addplot [ultra thick, red, forget plot]
table {%
3.9 2
4.1 2
};
\addplot [ultra thick, forestgreen4416044, dotted]
table {%
3.9 2.78333333333333
4.1 2.78333333333333
};
\addlegendentry{Mean}
\end{groupplot}

\end{tikzpicture}
\caption{Subjective evaluation results across naturalness and speaker similarity for the low-resource speaker.}
\label{fig:subj}
\end{figure*}
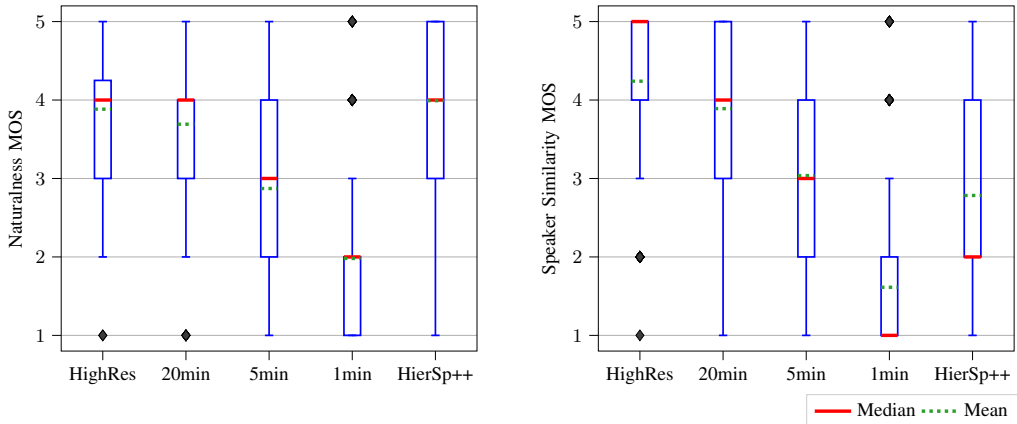

We use Mel Cepstral Distortion with Dynamic Time Warping (MCD-DTW) \cite{kayyar2023_subj_eval_ssw} as the objective measure to evaluate the synthesis quality. A lower MCD-DTW indicates that the synthesized samples are closer to the ground truth. The similarity of synthesized speech to the target voice was measured by computing the cosine similarity between the ground-truth and synthesized samples w.r.t.\ ECAPA-TDNN\cite{brecht2020_interspeech_ecapa} embeddings generated by SpeechBrain \cite{ravenelli2021_arxiv_speechbrain}. A higher cosine similarity value indicates that the samples are closer to the target.

The objective metrics for the LR speaker shown in Table~\ref{tab:eval-metrics} are averaged over 100 unseen sentences from each LR speaker dataset. The first row has the metrics for the original ForwardTacotron model trained using five HR speakers, providing an upper limit for the performance of our method. The next three rows correspond to our main experiments. We compared our experiments with two different baselines (HierSpeech++ and AdapterMix) and added ablations as the last part of the table. 

Across all test cases and objective metrics, our approach using 20 minutes of LR speaker data matched the performance of the ForwardTacotron HR training most closely. Specifically, the speaker similarity, assessed via cosine similarity, differs by just 0.09 (TC-Star) or 0.08 (Hi-Fi-TTS) between the model trained on over 6 hours (23 hours for Hi-Fi-TTS) of data and our proposed approach trained on only 20 minutes of data. 

The HierSpeech++ baseline has MCD scores worse than our model trained with 5 minutes of LR speaker data across both LR speakers. The MCD score for the female speaker obtained with HierSpeech++ is even higher (i.e., worse) than the one obtained with our version trained with one minute of data. Regarding speaker similarity, our model trained with only one minute of LR speaker data yields cosine similarity values that are 0.04 and 0.15 higher than the ones from HierSpeech++ for male and female speakers, respectively. The lower speaker similarity also explains the inferior MCD of HierSpeech++, which deteriorates when the target speakers voice timbre is not captured well. 

When using the same 20-minute LR subset for training, the objective metrics obtained with AdapterMix are worse than those obtained with our proposed approach.  

Although the AdapterMix outputs had better metrics than the HierSpeech++ outputs, the AdapterMix outputs exhibited artifacts from their vocoder model and were hence not used in the subjective tests. The results of all the models are shared in the accompanying website: \url{https://s.fhg.de/ftlrtts}.

 
 \subsection{Subjective Evaluation}
 \label{subsec:subj}

Complementing the objective evaluation, we used Mean Opinion Score (MOS) tests to evaluate the synthesized speech's naturalness and similarity to the target voice. These tests were realized with the webMUSHRA software \cite{SchoefflerEtAl15_webMUSHRA_WAC}. We conducted the test for naturalness following the ITU P.808 recommendation \cite{itu_p808_2018} to obtain a naturalness MOS.
 
 For the speaker similarity, we used a test similar to the one performed as a part of Blizzard Challenge 2023 \cite{perrotin23_csl_blizzard}. We presented one unseen ground-truth sample and one sample from each model under test per page of the listening test. The text content of the ground-truth samples differed from that of the test samples, which presented the same text across the different models. The listeners were instructed to assign each of the test samples to one of the following categories: 1 (completely different person), 2 (probably different person), 3 (similar), 4 (probably the same person), and 5 (exactly the same person). They had to listen and assign all the samples before moving to the next test page. The order in which the samples were presented was randomized. The scores across the samples and listeners were averaged to get the speaker similarity MOS, with a higher score indicating that synthesized speech was closer to the ground truth. 

 Both listening tests were done independently using twenty samples from the Harvard sentences \cite{rothauser69_Harvard} using the male speaker only. For the naturalness test, we had eighteen listeners with no known hearing defects and an average age of 31.4. For the speaker similarity test, we had fifteen listeners with no known hearing defects and an average age of 32.1.    

 Boxplots for the distributions of subjective ratings for the naturalness and speaker similarity MOS tests are visualized in Fig.~\ref{fig:subj}. The HR baseline had the best ratings, as expected. HierSpeech++ and our version using a 20-minute LR subset had similar results in terms of naturalness. Our versions using less data had lower naturalness ratings. Regarding speaker similarity, our proposed models trained with 20 and 5 minutes of data were rated 1.1 and 0.25 MOS points above the HierSpeech++ baseline, respectively. These listening test results indicate that, even though few-shot approaches like HierSpeech++ can synthesize highly natural speech, our approach of using little training data better captures the speaker identity of the target LR speaker.

\subsection{Discussion}
\label{subsec:ablation}

We conducted an ablation study using five minutes of target LR speaker data for completeness. The results of this study are presented in Table~\ref{tab:eval-metrics}. When noise augmentation is used but no binned sampling, the MCD-DTW value worsens by 7.6~dB and cosine similarity by 0.16 for TC-Star. Additionally, deactivating noise augmentation does not lead to any further decrease in MCD-DTW and results in a minimal decline (0.03) in cosine similarity. The trends differ for Hi-Fi-TTS. In this case, binned sampling does not affect the results. When noise augmentation and binned sampling are omitted, there is a 3.9~dB increase (worsening) in MCD-DTW and a decrease (worsening) of 0.21 in cosine similarity. This ablation study shows that noise augmentation and binned sampling affect different datasets differently, and a combination of both techniques is needed to obtain consistent results.

In addition to the previous experiments, we verified our approach's stability by training three acoustic models using separate 5-minute subsets containing different short sentences. When objectively evaluating these models, we obtained a slight difference of 0.4~dB in MCD-DTW across these models. The corresponding cosine similarity difference was 0.015. To test if the availability of short samples is required for our approach, we generated an LR set by splitting long sentences at speech pauses using WhisperX \cite{bain2023_whisperx_interspeech} instead of using genuine short samples from the dataset. When training an acoustic model on this data, the model yielded similar MCD-DTW and cosine similarity values compared to the model trained on short samples. Informal listening tests showed a substantially decreased synthesis quality when long sentences without splitting were used for training.

\section{Conclusion}
\label{sec:end}
We have proposed an approach for synthesizing natural-sounding, high-quality speech using limited training samples for the target speaker using noise augmentation and binning techniques. With only twenty minutes of data, we achieve a naturalness similar to ForwardTacotron trained with more than five hours of data.  Although the approach has been demonstrated using English training sets, it can also be applied to other languages for which only little data is available.

We compared our proposed techniques to two models: the zero-shot HierSpeech++ and the finetuning-based AdapterMix. HierSpeech++ requires more than 100 high-resource speakers during training but only needs a few seconds of target speaker data without retraining. In contrast, the proposed approach, which involves retraining, uses data from four high-resource speakers and 20 minutes of the target low-resource speaker data. This method improves speaker similarity by 1.1 MOS points but slightly reduces naturalness by 0.2 MOS points compared to HierSpeech++. Also, the proposed model is computationally efficient and has less than half as many parameters as HierSpeech++. Although AdapterMix is faster to retrain/finetune than the proposed model using the same amount of LR data, it delivers lower naturalness and speaker similarity than the proposed method.


\filbreak
 

\bibliographystyle{IEEEtran}
\bibliography{tts_references}

\end{document}